\documentclass{article}
\usepackage{amssymb}
\usepackage{amsmath}

\begin{document}
\centerline{\Large \bf The structure of general solutions}
\medskip
\centerline{\Large \bf and integrability conditions}
\medskip
\centerline{\Large \bf for rational first-order ODE's}

\vskip 2cm
\centerline{\sc Yu. N. Kosovtsov}
\medskip
\centerline{79060, 43-52, Nautchnaya Street, Lviv, Ukraine.}
\centerline{email: {\tt kosovtsov@escort.lviv.net}}
\vskip 1cm
\begin{abstract}
In present paper we propose an approach based on examination of the structure of the general solution of equations of the type $dy/dx=P(x,y)/Q(x,y)$, with $P$ and $Q$ polynomials only in $y$. Under the term structure we mean the dependency character of solution from arbitrary constant. We describe a common form of the structures for foregoing equations. In such a way one can obtain a differential-algebraic polynomial system for undetermined parameters of the structures. The successful solution of this system automatically leads to finding the general solution of ODE's. We demonstrate on examples that proposed method gives, as a first step, the systematic way for obtaining new integrability conditions and general solutions for various families of rational first-order ODE's.
\end{abstract}

\section{Introduction}

Equations of the type $dy/dx=P(x,y)/Q(x,y)$, with $P$ and $Q$ polynomials in $y$ arise in many different practical fields. There are some methods for finding the solutions for such type of equations. First of all it is the famous method of integrating factor. A remarkable method based on the knowledge of the general structure of the integrating factor was developed by Prelle and Singer \cite {Singer} for the case when $P$ and $Q$ are polynomials in \emph {both} arguments $x$ and $y$ (see also \cite {Duarte}).

In present paper we propose an approach based on examination of the \emph {structure} of the general solution of equations with $P$ and $Q$ polynomials \emph {only} in $y$. Under the term \emph {structure} we mean the dependency character of solution from \emph {arbitrary constant}. In theory of differential equations (especially linear) the use of solution structure is very fruitful. Such type of approaches allows us, for example, to obtain the general solution from the knowledge of definite number of particular solutions (superposition) or to use the method of undetermined parameters and so on.

Below we will see that many of practically important equations have sufficiently easy structures. The knowledge of the structure allows us to put up different kinds of solving strategies. In present paper we expose the preliminary results, which are directed on demonstration of the method abilities.

The paper is organized as follows: in section 2 we consider the base of the approach and discuss some possible solving strategies. In section 3 the well-known example of structure of Riccati equation is discussed from standpoint of proposed method. Sections 4-6 are devoted to specific examples, which expose some method features.

\section{Base of the approach}

The general solution of the first-order ODE
\begin{equation}
\frac{dy}{dx}=f(x,y)
\label{ode}
\end{equation}
is the function $y(x,c)$ of two variables $x$ and $c$, where $c$ is an "arbitrary constant".

Let is given a function of two variables $y(x,c)$. Is there an ODE of type (\ref{ode}) which general solution is given $y(x,c)$? To answer this question it is convenient to use well-known connection between first-order ODE's and linear first-order PDE's. If $\zeta (x,y)$ is a solution of linear PDE
\begin{equation}
\frac{\partial \zeta (x,y)}{\partial x}+ f(x,y)\frac{\partial \zeta (x,y)}{\partial y}=0,
\label{pde}
\end{equation}
then the general solution (in implicit form) of the equation (\ref{ode}) is
\begin{equation}
\zeta (x,y)=c,
\label{solode}
\end{equation}
and it being known that
\begin{equation}
y(x,\zeta )=c.
\label{solode2}
\end{equation}
So, if for given function $y(x,c)$ we solve algebraical equation (\ref{solode2}) in respect to $\zeta$ then from  (\ref{pde}) (under obvious restrictions) we obtain that the equation (\ref{ode}) with
\begin{equation}
f(x,y)=-\frac{\frac{\partial \zeta (x,y)}{\partial x}}{\frac{\partial \zeta (x,y)}{\partial y}}
\label{f}
\end{equation}
has as the general solution exactly the given function $y(x,c)$.

The examination of the expression (\ref{f}) shows that if
\begin{equation}
\zeta (x,y)=F[\frac{p_1 (x,y)}{q_1 (x,y)}+\alpha \ln \frac{p_2 (x,y)}{q_2 (x,y)}+\beta \arctan \frac{p_3 (x,y)}{q_3 (x,y)} ],
\label{form}
\end{equation}
where $F$ is an arbitrary differentiable function;  $p_i$, $q_i$ ($i=1,\dots,3$) are polynomials in $y$; $\alpha$ and $\beta$ are constants, then $f(x,y)$ is \emph {rational function in $y$}. If we seek solutions for equations with complex $f(x,y)$, then presence of $\arctan$ term is needless. The corresponding expression for $f(x,y)$ here is as follows ($\alpha=\beta=1$ for short):
\begin{align}
f(x,y) =&-[q_1(x,y)^2\frac {dp_2(x,y)}{dx}q_2(x,y)q_3(x,y)^2+\notag \\ &+q_1(x,y)^2\frac {dp_2(x,y)}{dx}q_2(x,y)p_3(x,y)^2-\notag \\ &-q_1(x,y)^2p_2(x,y)\frac {dq_2(x,y)}{dx}q_3(x,y)^2-\notag \\ &-q_1(x,y)^2p_2(x,y)\frac {q_2(x,y)}{dx}p_3(x,y)^2+\notag \\ &+\frac {dp_1(x,y)}{dx}q_2(x,y)p_2(x,y)q_1(x,y)q_3(x,y)^2+\notag \\ &+\frac {p_1(x,y)}{dx}q_2(x,y)p_2(x,y)q_1(x,y)p_3(x,y)^2-\notag \\ &-p_1(x,y)\frac {dq_1(x,y)}{dx}q_2(x,y)p_2(x,y)q_3(x,y)^2-\notag \\ &-p_1(x,y)\frac {dq_1(x,y)}{dx}q_2(x,y)p_2(x,y)p_3(x,y)^2+\notag \\ &+q_2(x,y)p_2(x,y)q_1(x,y)^2\frac {dp_3(x,y)}{dx}q_3(x,y)-\notag \\ &-q_2(x,y)p_2(x,y)q_1(x,y)^2p_3(x,y)\frac {q_3(x,y)}{dx}]/\notag \\ &[q_1(x,y)^2\frac {dp_2(x,y)}{dy}q_2(x,y)q_3(x,y)^2+\notag \\ &+q_1(x,y)^2\frac {dp_2(x,y)}{dy}q_2(x,y)p_3(x,y)^2-\notag \\ &-q_1(x,y)^2p_2(x,y)\frac {dq_2(x,y)}{dy}q_3(x,y)^2-\notag \\ &-q_1(x,y)^2p_2(x,y)\frac {dq_2(x,y)}{dy}p_3(x,y)^2+\notag \\ &+\frac {dp_1(x,y)}{dy}q_2(x,y)p_2(x,y)q_1(x,y)q_3(x,y)^2+\notag \\ &+\frac {dp_1(x,y)}{dy}q_2(x,y)p_2(x,y)q_1(x,y)p_3(x,y)^2-\notag \\ &-p_1(x,y)\frac {dq_1(x,y)}{dy}q_2(x,y)p_2(x,y)q_3(x,y)^2-\notag \\ &-p_1(x,y)\frac {dq_1(x,y)}{dy}q_2(x,y)p_2(x,y)p_3(x,y)^2+\notag \\ &+q_2(x,y)p_2(x,y)q_1(x,y)^2\frac {dp_3(x,y)}{dy}q_3(x,y)-\notag \\ &-q_2(x,y)p_2(x,y)q_1(x,y)^2p_3(x,y)\frac {dq_3(x,y)}{dy}]
\label{genform}
\end{align}

We can interpret the relation (\ref{genform}) from different points of view. If we, for example, specify definite expressions for $p_i$ and $q_i$, we can obtain a family (depending on parameters of the polynomials) of first-order ODE's with known (predetermined) structure of general solution which in implicit form is given by (\ref{solode}). Such interpretation as a first step of analysis is very essential in methodical sense.

Practically more important is an attempt to solve inverse problem: guided by some consideration of orders of polynomials we can try to equate given rational function $f(x,y)$ with a right-hand of (\ref{genform}) with undetermined functions - coefficients of polynomials. In such process when we equate polynomial coefficients we obtain a differential-algebraic \emph {polynomial} system of ODE's for undetermined parameters. The successful solution of this system (we have to stress here that it is sufficient to find \emph {particular} solutions) automatically leads to finding the general solution of ODE of type (\ref{ode}) with given rational function $f(x,y)$ as right-hand term. Since in intermediate stage a \emph {polynomial} system arises, there is hope that we can handle (simplify) this system by existent CAS elimination algorithms, e.g., by using \emph {casesplit} from \emph {ODEtools} in \emph {Maple}.

There are other possible variants of solving strategies, as e.g., on that way it is easy to reveal the structure of integrating factor and further to seek solution by means of undetermined parameters.

At last under known structure of solution (in such a way it is sufficient to ascertain the fact of existence of a solution of the system for undetermined parameters with given $f(x,y)$) we can try to find the necessary number of particular solutions of initial ODE and further to evaluate undetermined parameters via the set of particular solutions (non-linear superposition).

Let us count roughly the orders in $y$ of numerator and denominator of  rational function $f(x,y)$ in (\ref{genform}). Let us denote orders of polynomials $p_i$ and $q_i$ as $n_{p_i}$ and $n_{q_i}$; $N=\sum _{i=1}^3 (n_{p_i}+n_{q_i})$. It is obvious, that order of polynomial $P$ in numerator
\begin{align}
n_P\leq \max \{&(N+n_{q_1}+n_{q_3}-n_{p_1}-n_{p_3}),\notag \\ &(N+n_{q_1}-n_{q_3}-n_{p_1}+n_{p_3}),\notag \\ &(N+n_{q_3}-n_{p_3}),\notag \\ &(N-n_{q_3}+n_{p_3}),\notag \\ &(N+n_{q_1}-n_{p_1})\}
\notag
\end{align}
and order of polynomial $Q$ in denominator
\begin{equation}
n_Q\leq n_P-1.
\notag
\end{equation}
The maximal possible number of parameters (the general number of coefficients of polynomials $p_i$ and $q_i$) in the right-hand of (\ref{genform}) is $N+6$ ($N+4$ when $\arctan$ term is absent, or $N+2$ when $\ln$ term is absent too). As every rational function may be reduced by a nonzero factor and by virtue of elementary identities of type $\ln(\alpha(x)\phi(x,y))=\ln\alpha(x)+\ln\phi(x,y)$ then the real number of free parameters is only $N+1$.

As we would like to obtain an arbitrary rational function assigned in advance which nominally would have $n_P+n_Q+2$ (or $n_P+n_Q+1$ free) parameters we conclude that generally speaking there are a deficit of free parameters. It means first of all that for the most cases the classical classification of ODE's on orders of polynomials \cite {Kamke} does not coincide with classification of structures of solutions. To one classical class of ODE's (e.g., Abel equations) corresponds the set of structures, depending of parameters of specific equation. The mechanism of structure selection (determination of degrees for polynomials $p_i$ and $q_i$) is similar to proposed in \cite {Singer}.

From another side, as a consequence of a deficit of free parameters,  above mentioned system can be solved if some definite conditions (integrability conditions or constraint) for functions-parameters of given equation are held.

In the next sections we demonstrate on examples that proposed method gives, as a first step, the systematic way for obtaining new integrability conditions and general solutions for various families of rational first-order ODE's.

\section{Riccati equations}

The structure of solution is known for only a few non-linear equations. The most prominent example is the Riccati equation, where
\begin{equation}
y(x,c)= - \frac {b0(x)\,c -a0(x)}{b1(x)\,c -a1(x)}
\label{formR2}
\end{equation}
or
\begin{equation}
\zeta (x,y)=\frac {a1(x)\,y + a0(x)}{b1(x)\,y + b0(x)}.
\label{formR}
\end{equation}

If we substitute (\ref{formR}) into (\ref{f}), we obtain that this structure corresponds to ODE's family with
\begin{align}
f(x,y)=&
\frac {\frac {da1(x)}{dx}b1(x)-a1(x)\frac {db1(x)}{dx}}{-a1(x)b0(x)+b1(x)a0(x
)}\,y^2+\notag \\&+\frac {\frac {da1(x)}{dx}b0(x)+\frac {da0(x)}{dx}b1(x)-a1(x)\frac {db0(x)}{dx}-a0(x)
\frac {db1(x)}{dx}}{-a1(x)b0(x)+b1(x)a0(x)}\,y+\notag \\&+\frac {\frac {da0(x)}{dx}b0(x)-a0(x)
\frac {db0(x)}{dx}}{-a1(x)b0(x)+b1(x)a0(x)}.
\label{fR}
\end{align}
The general form of the Riccati equations is
\begin{equation}
\frac{dy}{dx}=X2(x)\,y^2+X1(x)\,y+X0(x)
\label{R}
\end{equation}
If we now equate right-hand sides of (\ref{fR}) and (\ref{R}) we obtain the following ODE's system:
\begin{align}
\{\frac {\frac {da1(x)}{dx}b1(x)-a1(x)\frac {db1(x)}{dx}}{-a1(x)b0(x)+b1(x)a0(x
)}&= X2(x);\notag \\ \frac {\frac {da1(x)}{dx}b0(x)+\frac {da0(x)}{dx}b1(x)-a1(x)\frac {db0(x)}{dx}-a0(x)
\frac {db1(x)}{dx}}{-a1(x)b0(x)+b1(x)a0(x)} &=X1(x);\notag \\ \frac {\frac {da0(x)}{dx}b0(x)-a0(x)
\frac {db0(x)}{dx}}{-a1(x)b0(x)+b1(x)a0(x)}&=X0(x).\}
\label{sysR}
\end{align}
We have to solve this system for $a1(x)$, $a0(x)$, $b1(x)$ and $b0(x)$. Let us simplify the system with help of differential elimination procedure, e.g., by using \emph {casesplit} from \emph {ODEtools} in \emph {Maple} (everywhere in this paper we use \emph {casesplit} with the default rankings). As a result we obtain some \emph {cases}, the most interesting of which here is
\begin{align}
[&a1(x) =
(-a0(x)X2(x)b0(x)-\frac {da0(x)}{dx}b1(x)+\notag \\&+X1(x)b1(x)a0(x)+a0(x)\frac {db1(x)}{dx})/(b1(x)X0(x))
\label{sol1R}
\end{align}
\begin{align}&
\frac {d^2a0(x)}{dx^2} =
(-b1(x)X0(x)a0(x)\frac {dX2(x)}{dx}b0(x)-2b1(x)X0(x)\frac {da0(x)}{dx}X2(x)
b0(x)+\notag \\&+b1(x)X0(x)a0(x)\frac {d^2b1(x)}{dx^2}-b1(x)X0(x)X1(x)\frac {db1(x)}{dx}a0(x)+\notag \\&+b1(x)^2X0(x)X1(x)\frac {da0(x)}{dx}+b1(x)^2X0(x)\frac {dX1(x)}{dx}a0
(x)-\notag \\&-2(\frac {db1(x)}{dx})^2X0(x)a0(x)+2X0(x)X2(x)b0(x)a0(x)\frac {db1(x)}{dx}
+\notag \\&+2\frac {db1(x)}{dx}X0(x)\frac {da0(x)}{dx}b1(x)-b1(x)\frac {dX0(x)}{dx}a0(x)\frac {db1(x)}{dx}+\notag \\&+b1(x)\frac {dX0(x)}{dx}a0(x)X2(x)b0(x)+b1(x)^2\frac {dX0(x)}{dx}\frac {da0(x)}{dx}-\notag \\&-b1(x)^2\frac {dX0(x)}{dx}X1(x)a0(x))/(X0(x)b1(x)^2),
\label{sol2R}
\end{align}
\begin{align}&
\frac {db0(x)}{dx} =
(-X0(x)b1(x)^2+b1(x)b0(x)X1(x)-\notag \\&-b0(x)^2X2(x)+b0(x)\frac {db1(x)}{dx})/b1(x)]
\label{sol3R}
\end{align}
\begin{equation}
where \,[b1(x) \neq 0, X0(x) \neq 0]
\label{sol4R}
\end{equation}
Here the number of undetermined parameters of the structure is superfluous, so there are not additional (integrability) conditions, except obvious restrictions (\ref{sol4R}). We can conclude that any Riccati equation has the structure in form of (\ref{formR}). To express the unknown parameters $a1(x)$, $a0(x)$, $b1(x)$ and $b0(x)$ via parameters of Riccati equation we have to solve among others almost the same equation (\ref{sol2R}) but here it is sufficient to find its particular solution.

\section{Abel A type equations}

Let now consider by the same way the following slightly more complicated structure
\begin{equation}
\zeta (x, y) = a1(x)\,y + b1(x) +\ln(a2(x)\,y + b2(x))
\label{formAA}
\end{equation}
which corresponds to ODE's with
\begin{align}
f(x,y) =&
-(\frac {da1(x)}{dx}a2(x)\,y^2 +(\frac {da2(x)}{dx}+\frac {db1(x)}{dx}a2(x)+\notag \\ &+\frac {da1(x)}{dx}b2(x))\,y+\frac {db2(x)}{dx}+\frac {db1(x)}{dx}b2(x))/\notag \\ &(a1(x)a2(x)\,y+a1(x)b2(x)+a2(x)).
\label{fAA}
\end{align}

This is sub-family of Abel A class of equations (see \cite {Kamke}):

\begin{equation}
\frac {dy}{dx} =\frac {X2(x)\,y^2+X1(x)\,y+X0(x)}{Y1(x)\,y+Y0(x)}.
\label{AA}
\end{equation}
Unknown parameters here are defined by the following system
\begin{align}
\{\frac {da1(x)}{dx}a2(x) &= -X2(x);\notag \\
\frac {da2(x)}{dx}+\frac {db1(x)}{dx}a2(x)+\frac {da1(x)}{dx}b2(x) &=-X1(x);\notag \\ \frac {db2(x)}{dx}+\frac {db1(x)}{dx}b2(x) &= -X0(x);\notag \\ a1(x)a2(x) &= Y1(x);\notag \\ a1(x)b2(x)+a2(x)& = Y0(x)\}.
\label{sysAA}
\end{align}
In contrast to the Riccati equations here solutions of the system not always exist. One of the cases here is as follows
\begin{align}
[&b2(x) =(-X0(x)^2Y1(x)^4+Y1(x)^3X0(x)X1(x)Y0(x)-\notag \\ &-Y1(x)^3Y0(x)X0(x)\frac {dY0(x)}{dx}+Y1(x)^2\frac {dY0(x)}{dx}Y0(x)^2X1(x)+\notag \\ &+Y1(x)^2Y0(x)^2X0(x)\frac {dY1(x)}{dx}-Y1(x)\frac {dY1(x)}{dx}Y0(x)^3X1(x)-\notag \\ &-Y1(x)X1(x)X2(x)Y0(x)^3-Y1(x)\frac {dY0(x)}{dx}X2(x)Y0(x)^3+\notag \\ &+X2(x)^2Y0(x)^4+\frac {dY1(x)}{dx}X2(x)Y0(x)^4)/\notag \\ &(-4Y1(x)^2\frac {dY0(x)}{dx}X2(x)Y0(x)+Y1(x)(\frac {dY1(x)}{dx})^2Y0(x)^2+\notag \\ &+4Y1(x)X2(x)^2Y0(x)^2-4Y1(x)^2X2(x)Y0(x)X1(x)+\notag \\ &+Y1(x)^3X1(x)^2-2Y1(x)^2\frac {dY0(x)}{dx}\frac {dY1(x)}{dx}Y0(x)+\notag \\ &+4Y1(x)X2(x)Y0(x)^2\frac {dY1(x)}{dx}-2Y1(x)^2\frac {dY1(x)}{dx}Y0(x)X1(x)+\notag \\ &+Y1(x)^3(\frac {dY0(x)}{dx})^2+2Y1(x)^3\frac {dY0(x)}{dx}X1(x));\label{sol1AA}
\end{align}
\begin{align}
&\frac {db1(x)}{dx} =(-Y1(x)X0(x)\frac {dY1(x)}{dx}+Y1(x)X1(x)\frac {dY0(x)}{dx}+\notag \\ &+Y1(x)X1(x)^2-2Y1(x)X0(x)X2(x)-X1(x)Y0(x)X2(x)-\notag \\ &-Y0(x)X2(x)\frac {dY0(x)}{dx})/\notag \\ &(Y1(x)^2X0(x)+Y0(x)^2X2(x)-Y1(x)X1(x)Y0(x));
\label{sol2AA}
\end{align}
\begin{align}
&a2(x) =(-Y1(x)^2X0(x)-Y0(x)^2X2(x)+Y1(x)X1(x)Y0(x))/\notag \\ &(-\frac {dY1(x)}{dx}Y0(x)+Y1(x)\frac {dY0(x)}{dx}+X1(x)Y1(x)-2X2(x)Y0(x));\label{sol3AA} \end{align}
\begin{align}
&a1(x) =(2X2(x)Y1(x)Y0(x)+Y1(x)\frac {dY1(x)}{dx}Y0(x)-\notag \\ &-Y1(x)^2X1(x)-Y1(x)^2\frac {dY0(x)}{dx})/\notag \\ &(Y1(x)^2X0(x)+Y0(x)^2X2(x)-Y1(x)X1(x)Y0(x));\label{sol4AA}
\end{align}
\begin{align}
&\frac {d^2Y0(x)}{dx^2} =(Y1(x)^2X1(x)Y0(x)^2\frac {dX2(x)}{dx}+Y1(x)Y0(x)^3\frac {dX2(x)}{dx}\frac {dY1(x)}{dx}+\notag \\ &+3Y1(x)Y0(x)^2X2(x)^2\frac {dY0(x)}{dx}+Y1(x)^3\frac {dX0(x)}{dx}\frac {dY1(x)}{dx}Y0(x)-\notag \\ &-Y1(x)^3X0(x)X2(x)\frac {dY0(x)}{dx}-Y1(x)^3X0(x)\frac {dY0(x)}{dx}\frac {dY1(x)}{dx}+\notag \\ &+Y1(x)^3\frac {dX1(x)}{dx}Y0(x)\frac {dY0(x)}{dx}-Y1(x)^2X1(x)^2Y0(x)\frac {dY1(x)}{dx}-\notag \\ &-Y1(x)^2\frac {dX1(x)}{dx}Y0(x)^2\frac {dY1(x)}{dx}-Y1(x)^2Y0(x)^2X2(x)\frac {dX1(x)}{dx}+\notag \\ &+Y1(x)^2X0(x)(\frac {dY1(x)}{dx})^2Y0(x)-Y1(x)^2Y0(x)^2\frac {dX2(x)}{dx}\frac {dY0(x)}{dx}-\notag \\ &-2Y1(x)^2Y0(x)X2(x)(\frac {dY0(x)}{dx})^2-2X2(x)^3Y0(x)^3+\notag \\ &+X2(x)X0(x)X1(x)Y1(x)^3+3X2(x)^2Y0(x)^2X1(x)Y1(x)-\notag \\ &-Y1(x)^2X2(x)X1(x)^2Y0(x)-2Y1(x)^2X2(x)^2X0(x)Y0(x)-\notag \\ &-Y1(x)^2X1(x)Y0(x)\frac {dY0(x)}{dx}\frac {dY1(x)}{dx}+Y1(x)^2X0(x)X2(x)\frac {dY1(x)}{dx}Y0(x)-\notag \\ &-3Y1(x)^2X1(x)Y0(x)X2(x)\frac {dY0(x)}{dx}+3Y1(x)Y0(x)^2X2(x)\frac {dY0(x)}{dx}\frac {dY1(x)}{dx}-\notag \\ &-2Y1(x)^3X0(x)\frac {dX2(x)}{dx}Y0(x)+2Y1(x)^3\frac {dX0(x)}{dx}X2(x)Y0(x)-\notag \\ &-Y1(x)Y0(x)^3X2(x)\frac {d^2Y1(x)}{dx^2}+3Y1(x)Y0(x)^2X2(x)X1(x)\frac {dY1(x)}{dx}-\notag \\ &-Y1(x)^3X0(x)\frac {d^2Y1(x)}{dx^2}Y0(x)+Y1(x)^2X1(x)Y0(x)^2\frac {d^2Y1(x)}{dx^2}-\notag \\ &-Y0(x)^3X2(x)(\frac {dY1(x)}{dx})^2-3Y0(x)^3X2(x)^2\frac {dY1(x)}{dx}+\notag \\ &+Y1(x)^3X1(x)^2\frac {dY0(x)}{dx}+Y1(x)^3X1(x)(\frac {dY0(x)}{dx})^2-\notag \\ &-Y1(x)^4\frac {dY0(x)}{dx}\frac {dX0(x)}{dx}-Y1(x)^4\frac {dX0(x)}{dx}X1(x)+Y1(x)^4X0(x)\frac {dX1(x)}{dx})/\notag \\ &(-Y1(x)^4X0(x)-Y1(x)^2Y0(x)^2X2(x)+Y1(x)^3X1(x)Y0(x))];\label{sol5AA}\end{align}
\begin{align}where \,&[Y1(x)\frac {dY0(x)}{dx}+X1(x)Y1(x)-2Y0(x)X2(x)-Y0(x)\frac {dY1(x)}{dx} \neq 0,\notag \\ & Y1(x) \neq 0,\notag \\ & Y1(x)^2X0(x)+Y0(x)^2X2(x)-Y1(x)X1(x)Y0(x)\neq 0].
\label{sol6AA}
\end{align}
So we have to consider the equation (\ref{sol5AA}) as an \emph {integrability condition} for this case. If for some Abel A type equation the integrability condition (\ref{sol5AA}) is held true, then the undetermined parameters of the solution structure (\ref{formAA}) are easily calculated (in quadratures!) from (\ref{sol1AA})-(\ref{sol4AA}) and the general solution immediately follows from (\ref{solode}). Note that in the case of Abel A type of equation we have five parameters, four of them here are free and fifth is defined by the integrability condition.

The existent Maple ODEtools implementations do not find the solutions for equations considered here and below.

\section{Unclassified example 1}

With the following structure
\begin{equation}
\zeta (x,y)=\frac{a2(x)\,y^2 + a1(x)\,y+a0(x)}{b2(x)\,y^2 + b1(x)\,y+b0(x)}
\label{formAC}
\end{equation}
the more wealthy family of ODE's is connected:
\begin{align}
&f(x,y)=
((\frac{da2(x)}{dx}b2(x)-a2(x)\frac{db2(x)}{dx})\,y^4+(-a2(x)\frac{db1(x)}{dx}-a1(x)\frac{db2(x)}{dx}+\notag \\&+\frac{da2(x)}{dx}b1(x)+\frac{da1(x)}{dx}b2(x))\,y^3+(\frac{da2(x)}{dx}b0(x)+\frac{da0(x)}{dx}b2(x)-\notag \\&-a1(x)\frac{db1(x)}{dx}+\frac{da1(x)}{dx}b1(x)-a2(x)\frac{db0(x)}{dx}-a0(x)\frac{db2(x)}{dx})\,y^2+\notag \\&+(-a1(x)\frac{db0(x)}{dx}-a0(x)\frac{db1(x)}{dx}+\frac{da1(x)}{dx}b0(x)+\frac{da0(x)}{dx}b1(x))\,y-\notag \\&-a0(x)\frac{db0(x)}{dx}+\frac{da0(x)}{dx}b0(x))/\notag \\&((a1(x)b2(x)-a2(x)b1(x))\,y^2+\notag \\&+(2
a0(x)b2(x)-2a2(x)b0(x))\,y+a0(x)b1(x)-a1(x)b0(x)).
\label{fAC}
\end{align}
In general it is unclassified equations of the following type
\begin{equation}
\frac {dy}{dx} =\frac {X4(x)\,y^4+X3(x)\,y^3+X2(x)\,y^2+X1(x)\,y+X0(x)}{Y2(x)\,y^2+Y1(x)\,y+Y0(x)}.
\label{AC}
\end{equation}

Unknown parameters here are defined by the following system
\begin{align}
\{\frac{da2(x)}{dx}b2(x)-a2(x)\frac{db2(x)}{dx} &= X4(x);\notag \\
-a2(x)\frac{db1(x)}{dx}-a1(x)\frac{db2(x)}{dx}+\frac{da2(x)}{dx}b1(x)+\frac{da1(x)}{dx}b2(x) &= X3(x);\notag \\
\frac{da2(x)}{dx}b0(x)+\frac{da0(x)}{dx}b2(x)-a1(x)\frac{db1(x)}{dx}+\frac{da1(x)}{dx}b1(x)-\notag \\-a2(x)\frac{db0(x)}{dx}-a0(x)\frac{db2(x)}{dx} &= X2(x);\notag \\
-a1(x)\frac{db0(x)}{dx}-a0(x)\frac{db1(x)}{dx}+\frac{da1(x)}{dx}b0(x)+\frac{da0(x)}{dx}b1(x) &= X1(x);\notag \\
-a0(x)\frac{db0(x)}{dx}+\frac{da0(x)}{dx}b0(x) &= X0(x);\notag \\
a1(x)b2(x)-a2(x)b1(x) &= Y2(x);\notag \\
2a0(x)b2(x)-2a2(x)b0(x) &= Y1(x);\notag \\
a0(x)b1(x)-a1(x)b0(x) &= Y0(x)\}.
\label{sysAC}
\end{align}

Here we examine one of simple cases which corresponds to the equation (\ref{AC}) with $X4(x)=0$. Another more complicated case we will consider in the next section.
\begin{equation}
[a1(x) = Y2(x)/b2(x),
\label{sol1AC}
\end{equation}
\begin{equation}
b1(x) =(2b0(x)Y2(x)+2b2(x)Y0(x))/Y1(x),
\label{sol2AC}
\end{equation}
\begin{equation}
a0(x) =Y1(x)/2\,b2(x),
\label{sol3AC}
\end{equation}
\begin{align}
\frac {db0(x)}{dx}& =
-1/2(4X0(x)b0(x)Y2(x)+4X0(x)Y0(x)b2(x)-\notag \\&-b0(x)Y1(x)X1(x)-b0(x)
Y1(x)\frac{dY0(x)}{dx})/(Y1(x)Y0(x)),
\label{sol4AC}
\end{align}
\begin{align}&
\frac{db2(x)}{dx}=
(-4Y2(x)b2(x)Y0(x)X1(x)+4Y2(x)b2(x)Y1(x)X0(x)+\notag \\&+4Y2(x)b2(x)Y0
(x)\frac{dY0(x)}{dx}-b2(x)Y1(x)^2\frac{dY0(x)}{dx}-\notag\\&
-8X3(x)b2(x)Y0(x)^2-b2(x)Y1(x)^2X1(x)+4b2(x)Y1(x)Y0(x)X2(x))/\notag\\&
(-2Y1(x)^2Y0(x)+8Y0(x)^2Y2(x))
\label{sol5AC}
\end{align}
\begin{equation}
a2(x)=0,
\label{sol6AC}
\end{equation}
\begin{align}&
\frac{dY2(x)}{dx}=
(-Y1(x)^2Y0(x)X3(x)-Y1(x)^2X1(x)Y2(x)-\notag \\&-Y1(x)^2\frac{dY0(x)}{dx}Y2(x)+
4Y1(x)Y2(x)^2X0(x)+4Y1(x)Y0(x)Y2(x)X2(x)-\notag \\&-4Y0(x)Y2(x)^2X1(x)+
4Y0(x)Y2(x)^2\frac{dY0(x)}{dx}-4Y2(x)Y0(x)^2X3(x))/\notag \\&(4Y0(x)^2Y2(x)-
Y1(x)^2Y0(x)),
\label{sol7AC}
\end{align}
\begin{align}
\frac{dY1(x)}{dx} &=
(-8X0(x)Y0(x)Y2(x)^2+4Y2(x)X0(x)Y1(x)^2+\notag \\&+4Y2(x)\frac{dY0(x)}{dx}Y1
(x)Y0(x)-Y1(x)^3X1(x)-Y1(x)^3\frac{dY0(x)}{dx}+\notag \\&+2Y0(x)Y1(x)^2X2(x)-4
X3(x)Y1(x)Y0(x)^2)/\notag \\&(4Y0(x)^2Y2(x)-Y1(x)^2Y0(x)),
\label{sol8AC}
\end{align}
\begin{equation}
X4(x) =0],
\label{sol9AC}
\end{equation}
\begin{equation}
where \,[b2(x) \neq 0, Y0(x) \neq 0, Y1(x) \neq 0, 4Y0(x)Y2(x)-Y1(x)^2 \neq 0]
\label{sol10AC}
\end{equation}

Here we have three equations (\ref{sol7AC})-(\ref{sol9AC}) as integrability conditions. The undetermined parameters of the solution structure (\ref{formAC}) are calculated (in quadratures too) from (\ref{sol1AC})-(\ref{sol6AC}) and the general solution follows from (\ref{solode}). Note that for this type of equations we have seven parameters, five of them here are free.

Of course, this type of equation is defined by the structure with $a2(x)=0$ (or $b2(x)=0$ - see the first equation of (\ref{sysAC})) but here we see that more simple cases are particular cases of more complicated structures. The main problem on this way is the limit of complication of structures, which is tackleable by existent differential elimination procedures.

\section{Unclassified example 2}

Examples in sections 4,5 lead to general solutions in quadratures. In some tackleable cases we can receive solutions with some special functions. As such example let us consider one of the cases of simplification of system for structure (\ref{formAC}) of previous section. The case is as follows:
\begin{align}
[&a1(x) =
(Y1(x)^2a2(x)\frac{dY0(x)}{dx}Y2(x)-2Y1(x)^2Y0(x)Y2(x)\frac{da2(x)}{dx}+\notag \\&+Y1(x)^2a2(x)X1(x)Y2(x)-4Y1(x)a2(x)X0(x)Y2(x)^2-\notag \\&-4Y0(x)^2a2(x)X4(x)Y1(x)-4Y1(x)Y2(x)Y0(x)a2(x)X2(x)+\notag \\&+8Y2(x)^2Y0(x)^2\frac{da2
(x)}{dx}+4Y2(x)^2Y0(x)a2(x)X1(x)+\notag \\&+8Y2(x)a2(x)X3(x)Y0(x)^2-4Y2(x)
^2Y0(x)a2(x)\frac{dY0(x)}{dx})/\notag \\&(8Y0(x)^2X4(x)Y2(x)-2Y0(x)X4(x)Y1(
x)^2),
\label{sol1UN}
\end{align}
\begin{align}
&b1(x) =
(4Y2(x)^2Y0(x)a2(x)b2(x)X1(x)+Y1(x)^2Y2(x)b2(x)a2(x)X1(x)+\notag \\&+Y1(x)^2Y2(x)a2(x)b2(x)\frac{dY0(x)}{x}+2Y2(x)Y0(x)X4(x)Y1(x)^2-\notag \\&-2Y1(x)^2Y2(x)Y0(x)b2(x)\frac{da2(x)}{dx}+8Y2(x)^2Y0(x)^2b2(x)\frac{da2(x
)}{dx}-\notag \\&-4Y1(x)Y0(x)^2X4(x)a2(x)b2(x)-4Y1(x)b2(x)Y2(x)^2a2(x)X0(
x)-\notag \\&-4Y1(x)Y2(x)Y0(x)b2(x)a2(x)X2(x)-8Y2(x)^2Y0(x)^2X4(x)-\notag \\&-4Y2(
x)^2Y0(x)a2(x)b2(x)\frac{dY0(x)}{dx}+8Y2(x)a2(x)b2(x)X3(x)Y0(x)^2
)/\notag \\&(-2Y0(x)a2(x)X4(x)Y1(x)^2+8Y0(x)^2a2(x)X4(x)Y2(x)),
\label{sol2UN}
\end{align}
\begin{align}
&a0(x) =
(Y1(x)^3a2(x)\frac{dY0(x)}{dx}+8Y2(x)Y0(x)^2\frac{da2(x)}{dx}Y1(x)+\notag \\&+4Y2(x)Y0(x)a2(x)Y1(x)X1(x)-4Y2(x)Y0(x)Y1(x)a2(x)\frac{dY0(x)}{dx}-\notag \\&-4Y
2(x)Y1(x)^2a2(x)X0(x)-16a2(x)X4(x)Y0(x)^3+\notag \\&+8a2(x)X3(x)Y1(x)Y0
(x)^2-4Y0(x)a2(x)Y1(x)^2X2(x)-\notag \\&-2Y0(x)Y1(x)^3\frac{da2(x)}{dx}+a2(x)
Y1(x)^3X1(x))/\notag \\&(16Y0(x)^2X4(x)Y2(x)-4Y0(x)X4(x)Y1(x)^2),
\label{sol3UN}
\end{align}
\begin{align}
&b0(x) =
(Y1(x)^3a2(x)b2(x)\frac{dY0(x)}{dx}-4Y2(x)Y0(x)Y1(x)a2(x)b2(x)\frac{dY0(x)}{dx}+\notag \\&+8Y2(x)Y0(x)^2b2(x)\frac{da2(x)}{dx}Y1(x)-8Y2(x)Y1(x)Y0(
x)^2X4(x)+\notag \\&+4Y2(x)Y0(x)a2(x)Y1(x)b2(x)X1(x)-4Y0(x)b2(x)a2(x)Y
1(x)^2X2(x)-\notag \\&-4Y2(x)Y1(x)^2b2(x)a2(x)X0(x)-16a2(x)b2(x)X4(x)Y0
(x)^3+\notag \\&+8a2(x)b2(x)X3(x)Y1(x)Y0(x)^2+2Y0(x)Y1(x)^3X4(x)-\notag \\&-2Y0(x)
b2(x)Y1(x)^3\frac{da2(x)}{dx}+b2(x)a2(x)Y1(x)^3X1(x))/\notag \\&(-4Y0(x)a2(x)
X4(x)Y1(x)^2+16Y0(x)^2a2(x)X4(x)Y2(x)),
\label{sol4UN}
\end{align}
\begin{equation}
\frac{db2(x)}{dx} =-(-\frac{da2(x)}{dx}b2(x)+X4(x))/a2(x),
\label{sol5UN}
\end{equation}
\begin{align}&
\frac{d^2a2(x)}{dx^2} =\notag \\&=
(8a2(x)Y0(x)X4(x)X0(x)X3(x)Y1(x)+8Y2(x)a2(x)Y0(x)^2X1(x)\frac{dX4(x)}{dx}-\notag \\&-2a2(x)Y0(x)X4(x)Y1(x)^2\frac{dX1(x)}{dx}-8a2(x)Y0(x)^2Y
1(x)X2(x)\frac{dX4(x)}{dx}+\notag \\&+8a2(x)Y0(x)^2Y1(x)X4(x)\frac{dX2(x)}{dx}+2a
2(x)Y0(x)Y1(x)^2\frac{dX4(x)}{dx}\frac{dY0(x)}{dx}+\notag \\&+2a2(x)Y0(x)Y1(x)^2\frac{dX4(x)}{dx}X1(x)+4Y2(x)a2(x)Y0(x)X1(x)^2X4(x)-\notag \\&-16a2(x)Y0(x)^2
X4(x)^2X0(x)-16a2(x)Y0(x)^3X4(x)\frac{dX3(x)}{dx}-\notag \\&-a2(x)X4(x)Y1(x)^2
X1(x)^2+a2(x)X4(x)(\frac{dY0(x)}{dx})^2Y1(x)^2-\notag \\&-16X4(x)X0(x)^2a2(x)Y2
(x)^2-16Y2(x)a2(x)Y0(x)X4(x)X0(x)X2(x)+\notag \\&+8Y2(x)a2(x)X4(x)X1(x)
X0(x)Y1(x)+16Y2(x)Y0(x)^3\frac{da2(x)}{dx}\frac{dX4(x)}{dx}-\notag \\&-4Y0(x)^2Y1
(x)^2\frac{dX4(x)}{dx}\frac{da2(x)}{dx}+16a2(x)Y0(x)^3X3(x)\frac{dX4(x)}{dx}-\notag \\&-8Y2(x)a2(x)Y0(x)^2X4(x)\frac{dX1(x)}{dx}+8Y2(x)a2(x)Y0(x)Y1(x)X
4(x)\frac{dX0(x)}{dx}-\notag \\&-4Y2(x)a2(x)Y0(x)X4(x)(\frac{dY0(x)}{dx})^2-8Y2(x)a2(x)Y0(x)^2\frac{dY0(x)}{dx}\frac{dX4(x)}{dx}-\notag \\&-2a2(x)Y0(x)X4(x)Y1(x)^2\frac{d^2Y0(x)}{dx^2}+8Y2(x)a2(x)Y0(x)^2X4(x)\frac{d^2Y0(x)}{dx^2}-\notag \\&-8Y2(x)a2(x)Y0(x)Y1(x)X0(x)\frac{dX4(x)}{dx})/\notag \\&(-4Y1(x)^2Y0(x)^2X4(x
)+16Y0(x)^3X4(x)Y2(x)),
\label{sol6UN}
\end{align}
\begin{align}&
\frac{dY2(x)}{dx} =
(-Y1(x)^2Y0(x)X3(x)-Y1(x)^2X1(x)Y2(x)-\notag \\&-Y1(x)^2\frac{dY0(x)}{dx}Y2(x)+
4Y1(x)Y0(x)Y2(x)X2(x)+4Y1(x)Y0(x)^2X4(x)+\notag \\&+4Y1(x)Y2(x)^2X0(x)-
4Y0(x)Y2(x)^2X1(x)+\notag \\&+4Y0(x)Y2(x)^2\frac{dY0(x)}{dx}-4Y2(x)Y0(x)^2X3
(x))/\notag \\&(4Y0(x)^2Y2(x)-Y1(x)^2Y0(x)),
\label{sol7UN}
\end{align}
\begin{align}
\frac{dY1(x)}{dx} &=
(8X0(x)Y0(x)Y2(x)^2-4Y2(x)\frac{dY0(x)}{dx}Y1(x)Y0(x)-\notag \\&-4Y2(x)X0(x)
Y1(x)^2-8X4(x)Y0(x)^3+4X3(x)Y1(x)Y0(x)^2-\notag \\&-2Y0(x)Y1(x)^2X2(x)+Y
1(x)^3X1(x)+Y1(x)^3\frac{dY0(x)}{dx})/\notag \\&(-4Y0(x)^2Y2(x)+Y1(x)^2Y0(x))],
\label{sol8UN}
\end{align}
\begin{align}
where \,[a2(x) \neq 0, X4(x) \neq 0, Y0(x) \neq 0, 4Y0(x)Y2(x)-Y1(x)^2 \neq 0]
\label{sol9UN}
\end{align}

The equations (\ref{sol7UN}), (\ref{sol8UN}) are integrability conditions for this case. But equations for undetermined parameters of the structure (\ref{formAC}) include ubiquitous linear second-order equation (\ref{sol6UN}). From restriction (\ref{sol9UN}) we see that $a2(x) \neq 0$, so the particular solution $a2(x)=0$ is not suitable in this case and in fact we have to solve (\ref{sol6UN}) by existent methods. It is obvious that under some conditions the solutions of the equation (\ref{sol6UN}) may involve special functions.

\section{Conclusions}

We presented some aspects in solving rational first-order ODE's by examination of the structures of general solutions. Examples in sections 4-6 demonstrated that proposed method gives as a first step the systematic way for obtaining new integrability conditions and general solutions for various families of rational first-order ODE's.

Besides first-order ODE the method is suited for solving non-linear equations of higher order too. The properties of its solutions expect further studies.

\end{document}